\documentclass[prl,aps,showpacs,twocolumn,floatfix]
{revtex4-1}
\usepackage{graphicx}
\usepackage{bm}
\usepackage{amsmath}
\usepackage{amssymb}

\begin{document}

\title{Profile of a Galactic Spherical Cloud of Self-Gravitating Fermions}

\author{B. G. Giraud}
\email{bertrand.giraud@cea.fr}

\author{R. Peschanski}
\email{robi.peschanski@cea.fr}

\affiliation{Institut de Physique Th\'eorique,
Centre d'Etudes Saclay, 91191 Gif-sur-Yvette, France}

\begin{abstract}
The field which binds a thermal fermionic cloud is defined as a Hartree integral
upon its density. In turn, the density results from the field via a Thomas-Fermi
occupation of the local phase space. This defines a complete theory of all
properties and observables for the cloud. As an application to dark matter
halos, comparisons with astronomic  data on dwarf spheroidal galaxies are
provided and discussed. Estimates of the  elementary fermion mass are obtained,
serving as a phase-space bound on fermionic dark matter.
\end{abstract}
\pacs{95.35.+d 95.30.Cq 98.52.Wz} 
\maketitle

\section{I Introduction}

While there is strong evidence for the existence of dark matter, the
absence of its direct detection triggers many questions, such as i) are we
talking about light or heavy particles? ii) are they fermions, such as,
for instance, sterile neutrinos, or a new kind of BSM fermions,
or are they some bosons? iii) are they self-binding into balls by gravitation
only or is there a contribution from some additional interaction between the
dark particles? iv) what could be the mass of a dark particle? and so on.

The present work deals with the questions of
self-binding and elementary mass, under very simple hypotheses, namely a
fermionic nature of the particles, in non relativistic thermal
equilibrium, and a self-consistent gravitational 
field. Recently, such a hypothesis has been explored for dark matter made 
of sub-$keV$ degenerate fermions \cite{Domcke:2014kla,Alexander:2016glq}
(see also \cite{Randall:2016bqw} which considers a degenerate
core surrounded by a thermal envelope). In the following we 
consider the case of quasi-degenerate fermions, meaning that the temperature
to be considered is low enough such that the quantum
effects have to be taken into account together with the temperature. For the
same reason we are led to consider fermions presently in a non-relativistic 
thermal state. The comparison with data is made for fermionic clouds of dark
matter supposed to be associated with the known Dwarf Spheroidal Galaxies (DSG)
in the vicinity of the Milky Way \cite{truc,muche}. 

More generally, and independently from  this precise
hypothesis, the determination of the properties of thermal fermionic 
galactic clouds proved to be useful to determine model-independent 
phase-space lower bounds on the mass of fermionic dark matter, starting 
with the pioneering Ref. \cite{Tremaine:1979we} which considered an initial
thermal relativistic stage of the fermion cloud. The bound is due to the
existence of a maximal phase-space density  which stays valid after further
collision-less and dissipation-less evolution,
leading to a quite model-independent lower mass bound for the fermion.  Various 
phase-space bounds related to this method have been discussed and evaluated by 
different ways since then, noticeably 
in Refs. \cite {Dalcanton:2000hn,Boyarsky:2008ju,Destri:2012yn,Shao:2012cg}.

In our scheme, since our 
calculations define a complete theory of all properties and observables for the 
cloud, depending on its temperature, the determination (or the lower bound) of
an allowed elementary fermion mass from the cloud finds a natural application
to the problem of the elementary fermion mass using DSG data.

We will here define the gravitational field by a simple Hartree convolution
integral between the basic Newton attraction and the dark matter density. In
turn, this density will result from the field according to a Thomas-Fermi
approximation. This allows a closing of the algebra into a self-consistent
equation for the field, or a self-consistent equation for the density. Such
equations are reminiscent of the Poisson equation and can be easily
solved numerically. We thus obtain a complete description of the profile of a
dark object, and of properties estimated by astronomic observations, such as
velocity dispersion and  radial extension. Our description is parametrized by
five parameters, namely Planck's constant $h$, the Newton gravitational
constant $G$, the elementary dark particle mass $m$, a temperature $T$ and a
scale parameter, that turns out to be a non-relativistic velocity $v_0$. Then
the simple structure of our theory provides a formula that calculates the
elementary mass $m$ in terms of the other parameters and of astronomic
observations. 
 
The equations obeyed by a self-gravitating fermionic cloud at a temperature $T$
are displayed in Section II. They provide a rich zoology of density profiles
and related observables, via both analytical and numerical approaches.
Section III is a numerical application of our results to astronomic cases. It
shows somewhat realistic numbers, on the galactic scale.  We will even find
a glimpse of an answer to the question of the elementary masses, leading to
an evaluation of a lower mass bound on a dark fermion, when combining the data
on the various DSG's.
Finally Sections IV and V offer a discussion and conclusion.

\section{II Basic equations and formal consequences}

Spherical symmetry is assumed throughout this paper. In a first part,
corresponding to degenerate fermions (T=0), we also assume (and indeed check)
that the dark matter cloud has a finite radius $R$. The dark particles, assumed 
to be spin $\frac{1}{2}$ fermions with mass $m$, are bound by 
their self-consistent potential field $\phi$.  Let $\vec r$ mean the 
position coordinates and set $r \equiv |\vec r|$. Then,
\begin{equation}
\phi(r)=-G \int d^3\vec{r'} \frac{\rho(r')} {|\vec r - \vec {r'}|}\,,
\label{Hartree}
\end{equation}
also reads,
\begin{equation}
\frac{1}{4\pi\, G}\ \phi(r)=-\frac 1r \int_0^r dr'\, r^{\prime\, 2} \rho(r') 
                        \ -      \int_r^R dr'\, r'\, \rho(r')\,.
\label{basic}
\end{equation}
Here, $G$ is the gravitational constant and $\rho(r')$ is the mass density 
of the dark matter inside the spherical shell contained between radii $r'$
and $r'+dr'$. The first term in the right hand side of Eq.(\ref{basic})
accounts for the  mass internal to the sphere with radius $r$. 
The second term accounts for the  mass in the ``corona'', between 
$r$ and $R$. 

At the surface, the field reaches the value,
$\phi(R)=-G M/R$, where $M$ is the total mass, 
$M \equiv M(R) =4\pi \int_0^R dr'\, r^{\prime\, 2} \rho(r')$. Above the 
surface, $\phi$ continues as, $\phi(r)=-G M/r$, but, in the following, we
shall be concerned with internal properties only, namely $0 \le r \le R$.

The centripetal force reads, upon derivating Eq.(\ref{basic}),
\begin{equation}
- \frac{d \phi}{dr}=- \frac{G M(r)}{r^2},\ \ M(r)=4 \pi \int_0^r dr'
r^{\prime\, 2} \rho(r').
\label{centripete}
\end{equation}
The field $\phi$, an increasing function of $r$, is negative $\forall r$.

It can be noticed, incidentally, that the definition of $\phi$ from 
Eq.(\ref{Hartree}) makes $\phi$ be a Hartree potential \cite{hartree}. In principle, 
one should also consider a Fock potential \cite{fock}, the result of exchange terms in 
the fermionic gravitational interaction. This Fock term is neglected here.

With the well known, Thomas-Fermi approximation \cite{thomas}, the fermionic population
of each volume element of this large system is assumed, at zero temperature,
to make a Fermi sphere in momentum space, with some radius $p_F(r)$. 
Accordingly, the mass density reads,
\begin{equation}
\rho(r)=m\, \frac{8 \pi}{3 h^3}\, [p_F(r)]^3,
\label{ThoFer}
\end{equation}
where $h$ is Planck's constant and $m$ is the dark matter particle mass. 
At position $r$ the Fermi momentum, $p_F(r)$, is defined from a maximal 
energy level, $m\, \mu$, available for occupation,
\begin{equation}
\frac{p_F^2(r)}{2m}+m\, \phi(r)=m\, \mu,
\label{momFer}
\end{equation}
where the product, $m\, \mu$, defines a chemical potential. That 
radius $R$, where $\phi$ equates $\mu$, induces $p_F(R)=0$, hence the ``local 
Fermi sphere'' shrinks to zero and the Thomas-Fermi approximation experiences
difficulties with a turning point. In practice for the present theory, the
density, $\rho(R)$, vanishes according to Eq.(\ref{ThoFer}) since then $p_F=0$.
This defines the surface as we see it. Accordingly, we set $\mu=-G M/R$.

Let us now consider a cloud in a thermal equilibrium, the likely
result of a cosmological period when dark matter was not yet decoupled.
For a finite temperature $T$ the Fermi sphere at position $r$ is partly 
filled only, and there are ``compensating'' fermions above the Fermi surface, 
both facts described by the well known Fermi occupation formula,
\begin{equation}
\rho=\frac{8 \pi m}{h^3} \int_0^{\infty} p^2 dp\, \frac{1}{ 1 + \exp\left\{
\left[\frac{p^2}{2m} + m \psi\right]/(kT) \right\} },
\label{occupat}
\end{equation}
with $k$ the Boltzmann constant and $\psi \equiv \phi-\mu$. We might still
define the surface by the condition,
\begin{equation}
\psi(R)=0\, ,
\label{brute}
\end{equation}
but, above that surface, namely when $\psi$ becomes positive, the formula,
Eq.(\ref{occupat}), allows for residual fermions. Their probability in momentum
space will become exponentially small, but a density tail occurs in coordinate
space. It is easy to verify that, actually, the obtained solutions tell that
$\psi(r)$ is of order $\log r$ and $\rho(r)$ of order $r^{-2}$, when
$r \rightarrow \infty$. Hence a divergence
occurs for the total mass, another concern with the validity of a
Thomas-Fermi approach if one uses it too brutally. For the sake of caution
and rigor, we shall present in the following, when necessary, two sets of
results at least, namely
{\it i)} a set where integrals upon $r$ obtain from a cut-off at $R$ as
defined by Eq.(\ref{brute}), then
{\it ii)} a set where integrals are extended until $1.5\, R$, for example,
or more, to show tail effects.
(One might also argue that a momentum cut-off is necessary to avoid relativistic
situations, but we will stay in the non-relativistic regime.)

It is now convenient to introduce a velocity $v_0$ such that 
$\psi_0 \equiv \psi(0)=-v_0^2/2$, then the scalings, $p=m\, v_0\, q$ and 
$\psi(r)=v_0^2\, \chi(r)/2,$ where $q$ and $\chi$ become dimensionless. 
Then $\chi_0 \equiv \chi(0) = -1$, $\chi(R)=0$ and Eq.(\ref{occupat}) reads,
\begin{equation}
\rho(r)=\frac{8 \pi\, m^4\, v_0^3}{h^3} \int_0^{\infty} \frac{q^2 dq}{ 1 + 
\exp\{\, [q^2 + \chi(r)]/\eta\,\} },
\label{occupaa}
\end{equation}
where $\eta = k\,T/(m\,v_0^2/2)$. In the following, we shall only consider
cases where $\eta < 1$, for the obvious reason that quantum effects might
be washed out if the thermal energy order of magnitude, $kT$, exceeds the
kinetic energy order of magnitude, $mv_0^2/2$.

The density also reads,
\begin{equation}
\rho(r)=-\frac{2\, (\pi\, \eta)^{\frac{3}{2}}\, m^4\, v_0^3}{h^3}\, 
Li\left[\frac{3}{2},-e^{-\chi(r)/\eta} \right],
\label{occupab}
\end{equation}
where $Li$ is the polylogarithm function. Define now, 
\begin{equation}
{\cal L}_3(\chi,\eta) \equiv 3/4\, \sqrt{\pi}\, \eta^{\frac{3}{2}}\,
Li\left[\frac{3}{2},-e^{-\chi/\eta}\right].
\end{equation}
Notice that,
\begin{equation}
\lim_{\eta \rightarrow 0} {\cal L}_3(\chi,\eta)=-(-\chi)^{\frac{3}{2}},
\end{equation}
making easier the transition with  the degenerate fermion case. Moreover we find
\begin{equation}
\rho(r)=-\frac{8\, \pi\,  m^4\, v_0^3}{3\, h^3}\, 
{\cal L}_3(\chi,\eta).
\label{occupac}
\end{equation}
The second of Eqs.(\ref{centripete}) then becomes, 
\begin{equation}
M(r)=-\frac{32\, \pi^2\, m^4\, v_0^3}{3\, h^3} 
\, \int_0^r dr'\, r^{\prime\, 2} {\cal L}_3\left[\chi(r'),\eta \right].
\label{mass}
\end{equation}
Since $d\phi/dr$ and $d\psi/dr$ are trivially equal, because $\mu$ is a 
constant, the first of Eqs.(\ref{centripete}) reads,
\begin{equation}
-\frac{d \chi}{dr}=
\frac{64\, \pi^2\, m^4\, v_0\, G}{3\, h^3\, r^2}\, 
\int_0^r dr' r^{\prime\, 2} {\cal L}_3\left[\chi(r'),\eta \right],
\end{equation}
where we again used the scaling relating $\psi$ and $\chi$. A further
derivative yields,
\begin{equation}
\frac{1}{r^2}\, \frac{d}{dr}\, r^2\, \frac{d\chi}{dr}=-K_0\, 
{\cal L}_3\left[\chi(r),\eta \right],
\label{Poisson}
\end{equation}
with 
\begin{equation}
K_0=\frac{64\, \pi^2\, m^4\, v_0\, G}{3\, h^3}.
\end{equation}
This Eq.(\ref{Poisson}) is actually the Poisson equation for the present 
problem. Define now a length scale $r_0$ by the condition, $r_0^2\, K_0=1$,
namely
\begin{equation}
r_0= \frac{\sqrt{3}\, h^{\frac{3}{2}}}{8\, \pi\, m^2\, \sqrt{v_0\, G}}.
\label{range}  
\end{equation} 
Then implement the scalings, $r= r_0\, x$, $R=r_0\, X$, and $\chi(r)=
\xi(r/r_0)$. This yields, 
\begin{equation}
\frac{1}{x^2}\, \frac{d}{dx}\, x^2\, \frac{d\xi}{dx} = 
- {\cal L}_3\left[\xi(x),\eta \right].
\label{reduced}
\end{equation}
This simplifies into, 
$\frac{1}{x^2} \frac{d}{dx} x^2 \frac{d\xi}{dx}=[-\xi(x)]^{\frac{3}{2}}$,
for $\eta=0$.

\begin{figure}
\scalebox{0.60}{\includegraphics*{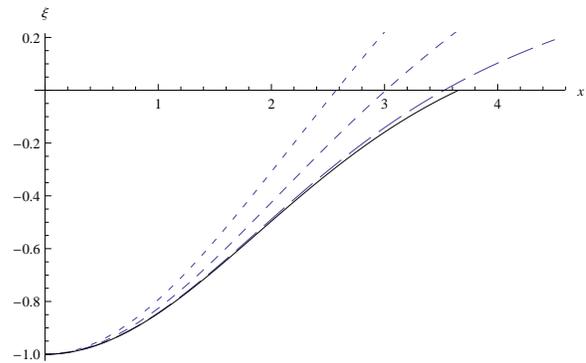}}
\caption{Scaled profiles $\xi$. Full line for $\eta=0$, long dashes for 
$\eta=.1$,  moderate dashes for $\eta=.3$,  short dashes for $\eta=.5$.}
\label{fieldforms}
\end{figure}

We show in Figure (\ref{fieldforms}) the solutions $\xi(x)$ numerically
obtained for $\eta=0,.1,.3.,.5$, respectively. We used the following boundary
conditions, $\xi(0)=-1$ and $d\xi/dx|_{x=0}=0$. The latter means that the field
has a smooth minimum at the center of the dark ball. 

\begin{figure}
\scalebox{0.60}{\includegraphics*{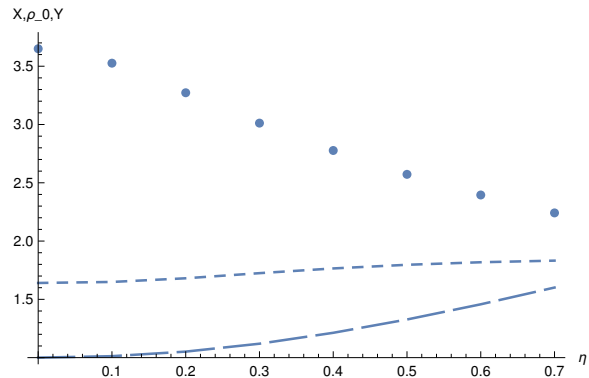}}
\caption{Evolutions in terms of $\eta$. Dots for the root $X$ of profile 
$\xi$, long dashes for the density $\rho(0)$ at center and short dashes
for the radius $Y$ at mid-density.}
\label{versustemp}
\end{figure}

\begin{figure}
\scalebox{0.60}{\includegraphics*{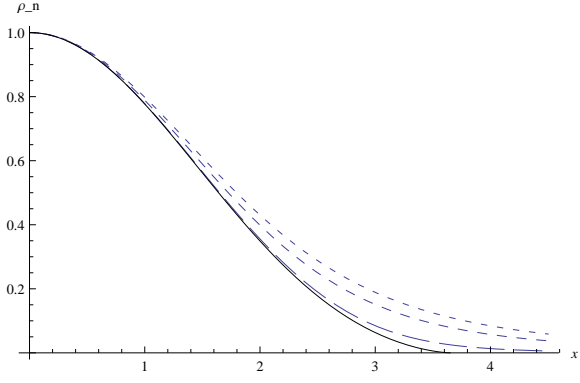}}
\caption{Renormalized densities $\rho_n(x)$. Full line, $\eta=0$.
Long, moderate, short dashes for $\eta=.1, .3, .5$. respectively.}
\label{normdens}
\end{figure}

Then we show, with a sequence of dots in Figure (\ref{versustemp}), the
evolution in terms of $\eta$ of the ``scaled radius'' $X$ where  $\xi$
vanishes. This is  of interest if the physical scale  of the ball radius is defined
as, $R=r_0 X$. It is seen that $X$ diminishes when $\eta$ increases. The same
Fig.(\ref{versustemp}) shows (long dashes) that the density at the origin
increases, and (short dashes) that the radius, $Y$, where the density is
half the density at center, $\rho(Y)=\rho(0)/2$, increases, but slightly
only. Notice that such conclusions, and most forthcoming ones, are valid only
for a fixed value of $v_0$ when $\eta$ changes. Cases with different values
of $v_0$ cannot be so trivially compared, since, given $m$, the present model
of a dark ball is driven by two independent physical parameters, namely $v_0$
and the temperature, $T=(mv_0^2/2) \eta/k$. Then $T$ occurs in $\eta$ only but
$v_0$ influences both $\eta$ and $r_0$.
In Figure (\ref{normdens}), we compare densities. The comparison is done
for normalized profiles, $\rho_n(x) \equiv \rho(x)/\rho(0)$. The full line
corresponds to $\eta=0$ and the long, moderate and short dashes to
$\eta=.1,.3,.5$, respectively. Tails grow when $\eta$ increases.

\begin{figure}
\scalebox{0.60}{\includegraphics*{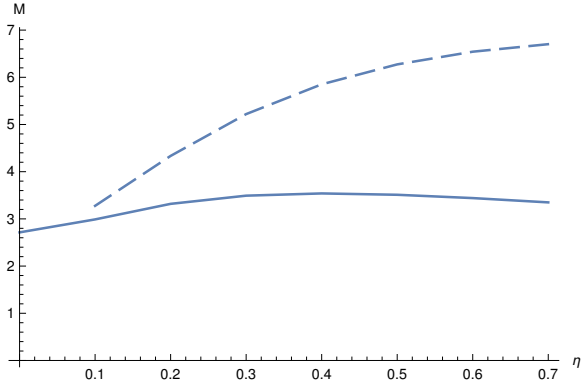}}
\caption{Total mass as a function of $\eta$. Full line for the mass interior to
the node of $\psi$, dashes for the mass including part of the tail.}
\label{depends}
\end{figure}

As is seen in Figure (\ref{depends}), the trend of the total mass is not
the same whether one calculates it with integrals limited to $X$ or extended
to $1.5\,X$. In the first case, $M$ first increases, then decreases as a
function of $\eta$. The second case cannot return a mass including tail
effects for $\eta=0$, since there is no tail. A systematic increase is then
found for finite values of $\eta$. 

\begin{figure}
\scalebox{0.60}{\includegraphics*{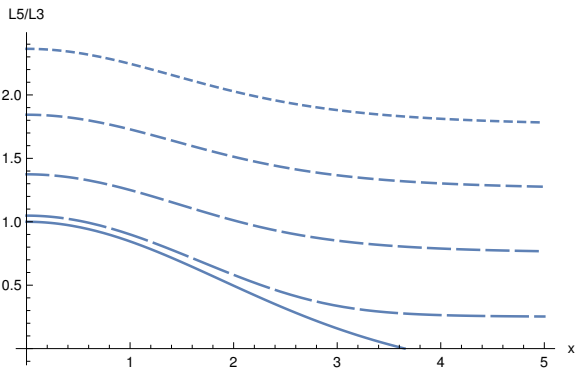}}
\caption{Profiles of mean square radial velocities. Full line, $\eta=0$.
Dashes shorten as $\eta=.1,.3,.5,.7$.}
\label{sqrad}
\end{figure}

Of special interest for our subject is the traditional observation \cite{tremaine} 
of the radial velocity dispersion for stars and the cosmological hypothesis
that the star dispersion is equal to that of the dark matter particles. 
Since in our Thomas-Fermi scheme the phase space density reads,
\begin{equation}
\rho_{ps}(\vec r,\vec p)\, d\vec r\, d\vec p = \frac{2\, h^{-3}\, d\vec r\, 
d\vec p} {1+\exp\{[\frac{p^2}{2m}+m\psi]/(kT)\}},
\end{equation}
the radial velocity dispersion for the dark particles is, inside a ball of
radius $R$,
\begin{equation}
\langle \frac{p^2}{3 m^2} \rangle =\frac{32 \pi^2}{3h^3m^2} 
\int_0^R \int_0^{\infty} 
\frac{r^2 dr\, p^4 dp}{1+\exp\{[\frac{p^2}{2m}+m\psi]/(kT)\}},
\label{dispers}
\end{equation}
namely, after now familiar scalings, $p=q\, m\, v_0$, $r=x\, r_0$, $R=X\, r_0$ 
and $\psi=\xi(r/r_0)\, v_0^2/2$,
\begin{equation}
\langle\frac{p^2}{3 m^2}\rangle= -{\cal K}\, v_0^2 \int_0^X x^2 dx\, 
Li\left\{ \frac{5}{2},-\exp \left[ \frac{-\xi(x)}{\eta} \right] \right\}.
\label{raddisp}
\end{equation}
with ${\cal K}=4\, (\pi\, \eta)^{\frac{5}{2}} (r_0\, m\, v_0/h)^3$.
Actually, it may be more interesting to use here the function,
\begin{equation}
{\cal L}_5(\xi,\eta) \equiv \frac{15\, \sqrt{\pi}\, \eta^{\frac{5}{2}}}{8} \,
Li\left[\frac{5}{2},-e^{-\frac{\xi}{\eta}}\right],
\end{equation}
with the advantage that
$\lim_{\eta \rightarrow 0} {\cal L}_5(\xi,\eta)=-(-\xi)^{\frac{5}{2}}$.
Then Eq.(\ref{raddisp}) becomes
\begin{equation}
\langle\frac{p^2}{3 m^2}\rangle= - \frac{32\, \pi^2 (m v_0 r_0)^3\, v_0^2}
{15\, h^3} \int_0^X x^2 dx\, {\cal L}_5[\xi(x),\eta].
\label{raddispsimpl}
\end{equation}
Note that the particle number reads,
\begin{equation}
N =\frac{32\, \pi^2}{h^3} \int_0^R \int_0^{\infty} 
\frac{r^2 dr\, p^2 dp}{1+\exp\{[\frac{p^2}{2m}+m\psi]/(kT)\}},
\label{number}
\end{equation}
or, as well, 
\begin{equation}
N = - \frac{32\, \pi^2 (m v_0 r_0)^3}{3\, h^3}
\int_0^X x^2 dx\, {\cal L}_3[\xi(x),\eta].
\label{numbersimpl}
\end{equation}
At $T=0$, Eq.(\ref{dispers}) reduces to,
\begin{equation}
\langle \frac{p^2}{3 m^2} \rangle =\frac{128 \sqrt{2}\, \pi^2 m^3 }{15\, h^3} 
\int_0^R r^2 dr\, \psi(r)^{\frac{5}{2}},
\end{equation}
hence, after scaling,
\begin{equation}
\langle \frac{p^2}{3 m^2} \rangle =\frac{32\, \pi^2 (m v_0 r_0)^3 v_0^2 }
{15\, h^3} \int_0^X x^2 dx\, \xi(x)^{\frac{5}{2}},\ X \simeq 3.6537.
\label{disprs0}
\end{equation}
Similarly, at $T=0,$
\begin{equation}
N =\frac{32\, \pi^2 (m v_0 r_0)^3 }{3\, h^3}  
\int_0^X x^2 dx\, \xi(x)^{\frac{3}{2}},\ X \simeq 3.6537.
\label{N0}
\end{equation}
For each value of $r$, or of the scaled radius $x$, we may define, from the
previous formulas, a local average radial dispersion, ${\cal L}_5/{\cal L}_3$,
where we discarded all inessential coefficients. We show in
Figure (\ref{sqrad}),
from top to bottom, the corresponding curves for $\eta=.7,.5,.3,.1$,
respectively. The length of dashes increases as $\eta$ diminishes.
The full line in Figure (\ref{sqrad}) shows the limit ratio, $-\xi$,
corresponding to $T=0.$ The line stops  beyond the sharp surface
radius, $X=3.6537$, naturally. Note how all finite temperature curves seem
to stabilize when $x$ exceeds that critical value, 3.6537. This might help
defining empirically  a ``surface'' for the ball, as a transition between a
``core'' and a ``tail'', typically at a radius of order, $R \sim 4 r_0$.
    
\begin{figure}
\scalebox{0.60}{\includegraphics*{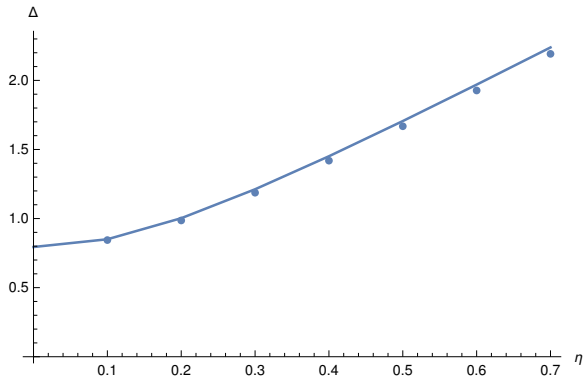}}
\caption{Mean square radial velocity in terms of $\eta$. Full line for integrals
interior to the node of $\phi-\mu$, dots for including part of the tail.}
\label{deltas}
\end{figure}

From Eq.(\ref{raddispsimpl}) and Eq.(\ref{numbersimpl}), the radial dispersion,
when averaged over the whole ball, reads
\begin{equation}
\Delta \equiv \sigma^2 \equiv \langle \frac{p^2}{3 m^2 } \rangle/N =
\frac{v_0^2}{5}\
\frac{\int_0^X x^2 dx {\cal L}_5(\xi,\eta)}
{\int_0^X x^2 dx {\cal L}_3(\xi,\eta)},
\label{Delt}
\end{equation}
where, for each $\eta$, we take $X$ as the root of $\xi$ or extend $X$ to 1.5
that root. For the case, $T=0$, ${\cal L}_5$ and ${\cal L}_3$ are replaced by
their now familiar limits and, obviously, there is no ``extended integral''
result. In Figure (\ref{deltas}), where the coefficient, $v_0^2/5$, is
voluntarily omitted, the results of this averaging of the radial
velocity dispersion over the ball volume are shown by a full line for integrals
limited by the field roots and by dots for ``extended integrals''. It will be
noted that tail effects seem to be small for the present observable. But, as
might be expected, the dispersion increases as a function of $\eta$.

\begin{figure}
\scalebox{0.60}{\includegraphics*{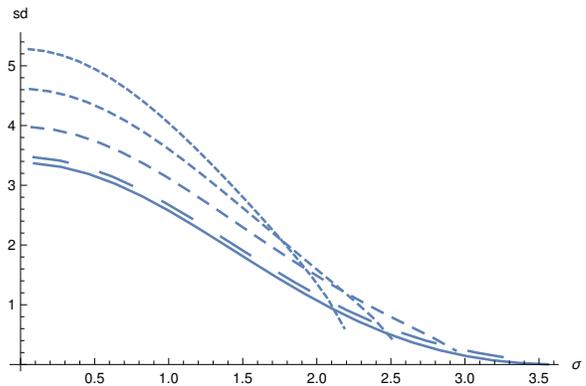}}
\caption{Surface densities when the radii are defined by the roots of $\xi$.
Full line, $\eta=0$. Dashes shorten as $\eta=.1,.3,.5,.7$.}
\label{surfdns1}
\end{figure}

\begin{figure}
\scalebox{0.60}{\includegraphics*{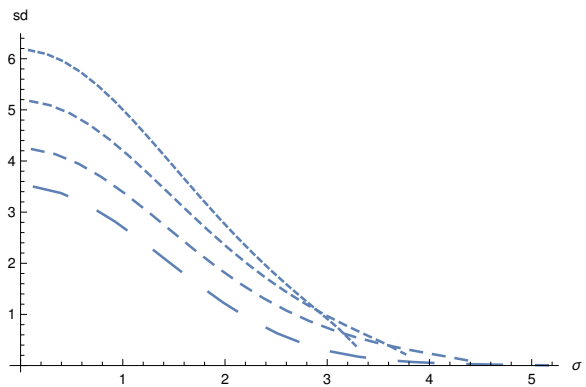}}
\caption{Surface densities with integration ranges increased by $50\%$.
Dashes shorten as $\eta=.1,.3,.5,.7$.}
\label{surfdns15}
\end{figure}

\begin{figure}
\scalebox{0.60}{\includegraphics*{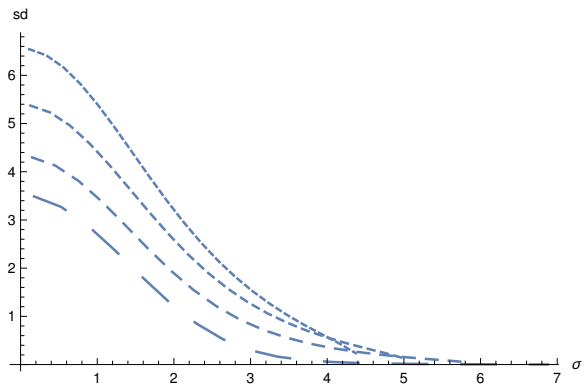}}
\caption{Same as Fig.(\ref{surfdns15}), but now with integration ranges twice
the roots of $\xi$.}
\label{surfdns2}
\end{figure}

\begin{figure}
\scalebox{0.60}{\includegraphics*{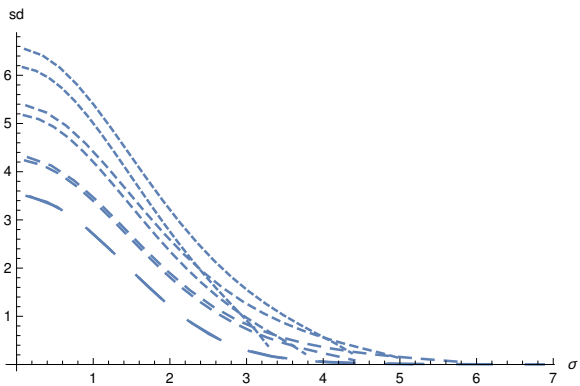}}
\caption{Comparison of the surface densities shown in Figs. (\ref{surfdns15})
and (\ref{surfdns2}).}
\label{fussurdns}
\end{figure}

Another concept used in galactic astronomy \cite{tremaine} is the surface density,
$\hat \rho(s)$, defined as follows. Consider an equatorial plane of the dark ball
and a (positive or negative) height $z$ measured from that plane. Coordinates inside
the plane may be chosen as polar coordinates, $\{s,\theta\}$ and the triplet,
$\{s,\theta,z\}$, make a simple set of cylindrical coordinates. Actually, the
angle, $\theta$, will be irrelevant in the following. The radius from the ball
center reads, obviously, $r=\sqrt{s^2+z^2}$. Given a radius $R$ for the ball,
the surface density is then defined as,
\begin{equation}
\hat \rho(s)=2 \int_0^{\sqrt{R^2-s^2}} dz\ \rho\left(\sqrt{s^2+z^2}\right).
\end{equation}
Upon taking advantage of Eq.(\ref{occupab}), this becomes, after an obvious
scaling of the form, $R=r_0 X$, $z=r_0 \zeta$ and $s=r_0 \sigma$,
\begin{equation}
\hat \rho(r_0 \sigma)=- \frac{8\, \pi\, m^4\, v_0^3\, r_0}{3\, h^3}
\int_0^{Z(\sigma)}
d\zeta\, {\cal L}_3\left[\xi\left(\sqrt{\sigma^2+\zeta^2}\right),\eta\right],
\end{equation}
where $Z(\sigma)=\sqrt{X^2-\sigma^2}$. Here again we shall compare situations
where $X$ is the root of $\xi$ or is increased by $50\%$ and even $100\%$.
The results are shown in Figures (\ref{surfdns1}), (\ref{surfdns15}) and
(\ref{surfdns2}). respectively. At low temperatures, a saturation of tail
effects is seen in Figure (\ref{fussurdns}), where Figs.(\ref{surfdns15})
and (\ref{surfdns2}) are fused.
    
A frequent observable used in astronomy is that radius, $r_c \equiv r_0 \tau$,
where the surface density reaches half the surface density at center of the
equatorial plane, namely,
\begin{equation}
\int_0^{Z(\tau)} d \zeta\
{\cal L}_3\left[\xi\left(\sqrt{\tau^2+\zeta^2}\right),\eta\right] =
\int_0^X d \zeta\ {\cal L}_3\left[\xi(\zeta),\eta\right]/2\, .
\label{midsurf}
\end{equation}
Clearly, this $r_c$ is not observed directly. Rather, it is estimated from
luminosity, decreasing from the center of the galaxy towards its edge. When
the luminosity has diminished by $50$\%, it is assumed that the same is true
for the underlying dark matter surface density.

\begin{figure}
\scalebox{0.60}{\includegraphics*{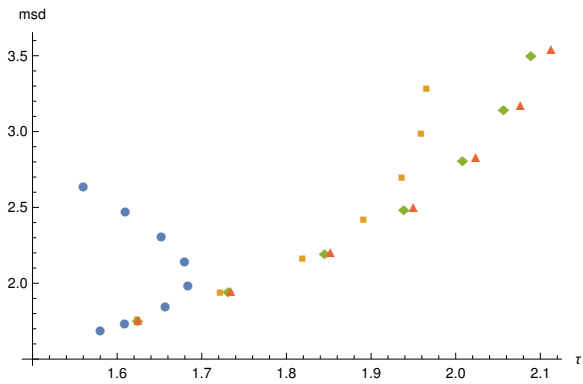}}
\caption{Evolutions of the halved surface density and its position.}
\label{midsur}
\end{figure}

We show in Figure (\ref{midsur}) four sets of solutions $\hat \rho(\tau;\eta)$
of Eq.(\ref{midsurf}). In all four sets, the mid-density increases as a
function of $\eta$. The first set, where the ranges of integrals are defined
by the nodes of the $\xi$'s, is illustrated by $8$ dots, corresponding to
$\eta=0,.1,.2,...,.7$, respectively. It indicates that $\tau$ first increases
then decreases. The second set results from integration ranges extended by
$100\%$. It contains $7$ points, with $\eta=.1,.2,...,.7$, and is illustrated
by squares. It hints again a lack of monotonicity for $\tau$, but
much less pronounced. The third set, illustrated by $7$ diamonds,
corresponds again to $\eta=.1,.2,...,.7$. It is obtained with ranges $5$ times
those of the first set and exhibits a monotonic trend for $\tau$. For the
fourth set, shown by triangles, with the same $7$ values of $\eta$, the range
extension factor is $10$. The set does not differ much from the previous one
and a saturation of tail effects can be expected. Notice how the four patterns
converge when $\eta$ diminishes. Actually, the lowest point of the first pattern
($\eta=0$, no extension) may be considered as belonging to all patterns as well.
We list here the $7$ values of $\tau$ obtained, when $\eta=.1,.2,\dots,.7$, and
the factor of integration range extension is $10$ :
$\{1.62,1.73,1.85,1.95,2.02,2.07,2.11\}$.

Now that we have at our disposal both quantities,
$\sigma^2=\langle p^2/(3 m^2 N) \rangle$ and $\tau$, we can return to realistic,
unscaled quantities and consider the ratio,
\begin{equation}
{\cal R}\equiv\frac{\sigma^2}{G\, \rho_0\, r_c^2} =\frac{\sigma^2}{G\, \rho_0\,
r_0^2\, \tau^2}
=\frac{8\pi\, \int {\cal L}_5}{5\, {\cal L}_3(-1,\eta)\, \tau^2\,
\int {\cal L}_3 }\, ,
\end{equation}
which has the advantage of being independent from $v_0$ and $m$. For $\eta=0$,
with integral ranges confined inside $R=3.6537\, r_0$ when we calculate
surface densities, we find, numerically, ${\cal R} \simeq 1.60$.
For $\eta=.1, .2,...,.7$ we find the following, respective values,
$1.59$, $1.55$, $1.53$, $1.52$, $1.51$, $1.51$, $1.51$.
These are obtained with integral ranges extended by a factor $10$, see the
sequence of triangles in Fig.(\ref{midsur}), because we want to take full
tail effects into account. It is obvious that all results are compatible with
a value ${\cal R} \simeq 1.55 \pm .05$. This privileges the King radius
\cite{tremaine}
found in the literature. Somewhat disappointing is the fact that our results
for ${\cal R}$ show little dependence on $\eta$ and, therefore, cannot suggest
a measure of the temperature of the dark matter cloud. But independence from
parameters, on the other hand, may lead to solid experimental testing.

Assume that the integration range $X$ is large enough to saturate tail effects.
Then, according to Eq.(\ref{Delt}),
$v_0^2=5\, \sigma^2\, {\cal F}(\eta)$, where, ${\cal F}$, in a schematic
notation, reads
${\cal F}(\eta)=\int {\cal L}_3\ /\int {\cal L}_5$. It is easy to tabulate.
As well,
according to Eq.(\ref{midsurf}), $\tau$ is a function ${\cal G}(\eta)$, for
which we listed seven values just a few lines above. Now, according to
Eq.(\ref{range}), we can invert logics and obtain a formula for the mass,
\begin{equation}
m^4 = \frac{3\, h^3} {64\, \pi^2\, v_0\, G\, r_0^2} = \frac{3\,h^3\,
[{\cal G}(\eta)]^2} {64\sqrt{5}\, \pi^2\, \sigma\,
\sqrt{{\cal F}(\eta)}\, G\, r_c^2}.
\label{masfrmul}
\end{equation}
With the amplification of the integration range taken as $10$, the list of
values for ${\cal F}(\eta)$, when $\eta=.1,.2,\dots,.7$, reads,
$\{ 1.19,1.02,.85,.71,.61,.53,.47 \}$.

With explicit numerical constants, Eq.(\ref{masfrmul}) also reads, in $eV$
units for $mc^2$, then $km/s$ for $\sigma$ and $pc$ (parsec) for $r_c$,
\begin{equation}
\left(\frac {mc^2}{100\ eV}\right)^4\! =\! 
\frac{96\cdot\{2.4, 3.0, 3.7, 4.5, 5.2, 5.9, 6.5 \}}
     {\sigma/(10\ kms^{-1})\, [r_c/(100\ pc)]^2}\, ,
\label{masfrmulb}
\end{equation}
where we use the typical observational scales.
The string in the numerator shows the $\eta$ dependent coefficient,
$[{\cal G}(\eta)]^2/\sqrt{{\cal F}(\eta)}$, with our usual set
$\eta=.1,.2,\dots,.7$.
When reduced to its $1/4$-th power, this coefficient does not seem to be very
influential.

\section{III Realistic illustrations}

Standard values in SI units for physical constants read,
$k=1.38\, 10^{-23}$, hence $k = 8.6\, 10^{-5}\, eV\, K^{-1}$,\ $h=6.6\ 10^{-34}$,
\ $G=6.7\ 10^{-11}$, \ $c=3.0\ 10^8$.
For the mass of dark matter particles, we shall first consider that a
tolerable guess is, $m \simeq 3.6\, 10^{-34}$, hence, $m c^2 \simeq 200\, eV$.
If $v_0 \simeq 10^{-4}\, c$, a non relativistic situation, then
$r_0 \simeq 6.4\, 10^{18}$, {\it i.e.}\, of order $210\, pc$. Recall that,
$X \simeq 3.5$ for $T=0$, and that, for positive temperatures, tails may more
than double the qualitative estimates of the radius. Then the cloud size may
reach the $kpc$ range. Incidentally, with such estimates for $m$ and
$v_0$, the coefficient, $\gamma \equiv .5\, m\, v_0^2 /k$, which converts
$\eta$ into a temperature $T$, is of order $\simeq 0.01$. We are here,
therefore, in a situation of very low temperatures if $\eta$ is kept $\leq 1$.

With different parametric assumptions, such as a lighter mass,
$m \simeq 2.0\, 10^{-34} \simeq 110\, eV/c^2$, and a higher reference velocity,
$v_0 \simeq 2\, 10^{-3} c$, we obtain $r_0 \simeq 4.6\, 10^{18}$, about
$150\, pc$,
and the temperature coefficient becomes, $\gamma \simeq 2.6$. This
value of $\gamma$ brings that range of $\eta$ values, studied in this paper,
able to induce temperatures much closer to the CMB temperature order
of magnitude.

From astronomic observations there might be some hope for measuring the
center density, $\rho_0 \equiv \rho(0)$. Typically,
$\rho_0 \sim 10^{-21}\ kg/m^3.$
From Eq.(\ref{ThoFer}) and again a guess of $m \simeq 110\ eV/c^2$, such a
value of $\rho_0$ would return,
\begin{equation}
p_0 \equiv p_F(0) \sim 5.6\, 10^{-23}\, (\rho_0)^\frac{1}{3} \, \sim 5.6\,
10^{-30}\, .
\end{equation}
This gives $v_0 \simeq 2.8\ 10^4\ m/s \simeq 9.\,  10^{-5} c$, lower than,
but not too far from the guesses made a few lines above.

More realistically, let us directly use Eq.(\ref{masfrmulb}) and
introduce orders of magnitude $\sigma=10^4$ m/s and $r_c=400\, pc$, as
suggested by astronomic observations. Then we obtain the string of estimates,
$mc^2=\{195, 206, 217, 228, 237, 244, 250\}\, eV$, according to $\eta$.

From \cite{truc}, we extract the following table of astronomic data and
corresponding results,

\ 

\begin{table}[h]
\centering
\begin{tabular}{|l|c|c|c|}
\hline
 Object    &      $r_c\ (pc)$   &    $\sigma\ (km/s)$  &  $mc^2\ (eV)$ \\
\hline
 Sextans   &     630 $\pm$ 170  &   6.6 $\pm$ 2.3  &  202 $\pm$ 32 \\
 Fornax    &     400 $\pm$ 103  &  10.5 $\pm$ 2.7  &  225 $\pm$ 32 \\
 LeoI      &     330 $\pm$ 106  &   8.8 $\pm$ 2.4  &  259 $\pm$ 45 \\
 UrsaMinor &     300 $\pm$  74  &   9.3 $\pm$ 2.8  &  268 $\pm$ 39 \\
 Carina    &     290 $\pm$  72  &   6.8 $\pm$ 1.6  &  295 $\pm$ 40 \\
 Draco     &     221 $\pm$  16  &   9.5 $\pm$ 1.6  &  311 $\pm$ 17 \\
 Bootes    &     246 $\pm$  28  &   6.5 $\pm$ 1.7  &  324 $\pm$ 28 \\
 Sculptor  &     160 $\pm$  40  &  10.1 $\pm$ 0.3  &  360 $\pm$ 45 \\
 Leo II    &     185 $\pm$  48  &   6.8 $\pm$ 0.7  &  369 $\pm$ 49 \\
\hline
\end{tabular}
\caption{Data from \cite{truc} and resulting elementary masses.}
\end{table}

\begin{figure}
\scalebox{0.60}{\includegraphics*{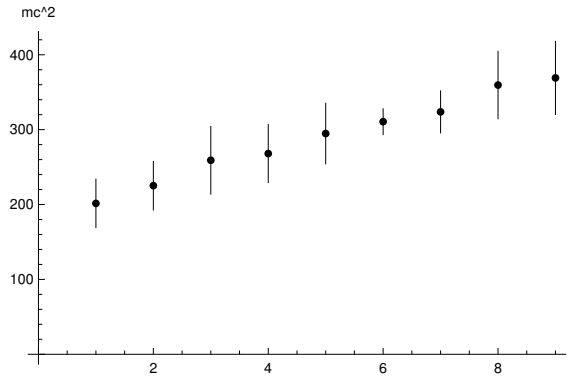}}
\caption{Masses and error bars deduced from the astronomic data in \cite{truc}.
Each integer abscissa means rank in Table I.}
\label{figtruc}
\end{figure}

For each astronomic object, we calculated the resulting elementary mass,
seen in the right-hand-side column of the table, from Eq.(\ref{masfrmulb}) with
the ``$\eta$ coefficient'' taken as $4.5$, an average value. Let $\Delta r_c$
and $\Delta \sigma$ be the absolute errors for $r_c$ and $\sigma$,
respectively. Then, according to Eq.(\ref{masfrmulb}), we estimated the
relative error for each estimated $m$ as,
$\Delta m/m=\left[\sqrt{4(\Delta r_c/r_c)^2+(\Delta \sigma/\sigma)^2}\right]/4$.
As seen in the table, this elementary mass spreads between $200$
and $370\ eV$, with error bars also spreading, between $\pm 30$ and
$\pm\ 50\ eV$. The pattern of values and error bars is illustrated by
Figure (\ref{figtruc}), their representation in increasing order being purely
a convention.

To compensate for this lack of precision, we implemented a $\chi^2$ fit for the 
obtained $m_i$, $m_i-\Delta m_i$ and $m_i+\Delta m_i$, $i=1,...,9$. This gives,
$mc^2=290 \pm 30\ eV$, with $\chi^2 = 2.5$ for the central value.

A similar set of data, taken from \cite{muche}, reads, with the resulting
estimates for the elementary mass again from Eq.(\ref{masfrmulb}) and again
the ``$\eta$ coefficient'' taken as $4.5$,

\

\begin{table}[h]
\centering
\begin{tabular}{|l|c|c|c|}
\hline
 Object        &  $r_c\ (pc)$   &   $v\ (km/s)$   & $mc^2\ (eV)$ \\
\hline
 CanVenat I    &  564 $\pm$  36 &  7.6 $\pm$ 2.2 & 206 $\pm$  16 \\
 UrsMaj I      &  318 $\pm$  45 &  7.6 $\pm$ 2.4 & 274 $\pm$  29 \\
 Hercules      &  330 $\pm$  63 &  5.1 $\pm$ 2.4 & 297 $\pm$  45 \\
 LeoT          &  178 $\pm$  39 &  7.5 $\pm$ 2.7 & 367 $\pm$  52 \\
 UrsMaj II     &  140 $\pm$  25 &  6.7 $\pm$ 2.6 & 426 $\pm$  56 \\
 LeoIV         &  116 $\pm$  30 &  3.3 $\pm$ 2.8 & 558 $\pm$ 139 \\
 ComaBeren     &   77 $\pm$  10 &  4.6 $\pm$ 2.3 & 630 $\pm$  89 \\
 CanesVenat II &   74 $\pm$  12 &  4.6 $\pm$ 2.4 & 644 $\pm$  99 \\
\hline
\end{tabular}
\caption{Data from \cite{muche} and resulting masses.}
\end{table}

\begin{figure}
\scalebox{0.60}{\includegraphics*{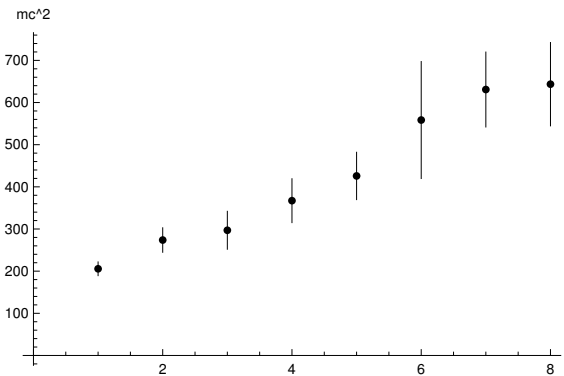}}
\caption{Masses and error bars, astronomic data from \cite{muche}.
Abscissa is rank in Table II.}
\label{figmuche}
\end{figure}

The estimates and error bars are shown in Figure (\ref{figmuche}). 
After the $\chi^2$ processing of such $8$ results, we obtain the following
estimate of the mass, $mc^2= 263 \pm 29\ eV$. It occurs with a disappointing
$\chi^2=8.9$, however. 

We stress, finally, that a compromise value, $mc^2 \simeq 275\ eV$,
reasonably sits in the overlap of the final error bars, $\pm\ 30\ eV$, observed
from the two distinct sets of data.

\section{IV Discussion}

Despite rather large uncertainties in astronomic  data on DSG's,
we have reached,
however, a reasonable estimate of the ``elementary dark mass'',
$mc^2 = 275 \pm 30\ eV $. This value, while a bit larger, is in a reasonable
agreement  with those obtained with different methods  in
Refs. \cite{Domcke:2014kla,Alexander:2016glq,Randall:2016bqw}, taking into
account the experimental error bars.

Concerning the role of temperature
in the range we consider, it may modify the resulting DSG-related masses 
by a maximum of an extra {15\%} amount only.

A few words of caution are in order, despite this agreement.
While our theory is consistent, we found that the dispersion
of obtained mass estimates between various clouds remains somewhat large,
with resulting large values of $\chi^2$'s. Our agreement, therefore, tends
to allow for the possibility that some DSG's could be  made of quasi
degenerate fermions, but it should not be taken as an absolute proof of such
a description for all DSG's. Another option is to turn to a different
interpretation, in terms of the determination of a lower fermionic mass bound.

Indeed, assuming a thermal equilibrium for an initial stage of the formation
of a dark matter cloud, then Dwarf Spheroidal Galaxies provide a  rather
model-independent, mass lower bound for a possible  elementary dark matter
fermion. Assume an initial thermal equilibrium state, like in Refs.
\cite{Dalcanton:2000hn,Boyarsky:2008ju,Destri:2012yn,Shao:2012cg} that we
referred to in the introduction, but with a temperature low enough to allow
neglecting relativistic corrections. Then let us list some distinct features
that we obtain. As discussed by the authors of
Ref. \cite{Boyarsky:2008ju},  the determination of a lower bound, from the
sequence of mass determinations for the DSG's of tables I and II, is
complicated by various dynamical indeterminacies. Following the choice made by
\cite{Boyarsky:2008ju}, and for the sake of a comparison,  we shall consider
the central value obtained from the DSG LeoIV as a typical lower bound.
This leads us to a mean value, $mc^2 \gtrsim 550\ eV $ (see Table II). This is
compatible with, while somewhat larger than, the result of the relativistic case
($mc^2 \gtrsim 480\ eV $) quoted in \cite{Boyarsky:2008ju}. Note that we also
find a consistency with the simulation-based determination,
$mc^2 gtrsim 500\, eV $, of 
Ref.\cite{Shao:2012cg}. Interestimgly, the agreement  is  even valid for each
of our individual mass determinations for DSG's in Tables I and II. All these
lie within 1 standard deviation from those quoted in \cite{Shao:2012cg}.

The results quoted in Refs.\cite{Dalcanton:2000hn,Destri:2012yn} are somewhat
different and larger, but a comparison is then not easy, because the  DSG's
which are considered by \cite{Dalcanton:2000hn,Destri:2012yn} are not the same
as those that we and Refs. \cite{Boyarsky:2008ju,Shao:2012cg} have used.

\section{V Conclusion}

The present theory allows the calculation of basically all properties of a dark
cloud, in terms of three parameters, namely an elementary particle mass $m$, a
velocity scale $v_0$ and a temperature scale $\eta$. It suffers from a
weak divergence due to density tail effects, but we have seen that many
properties are not sensitive to the divergence. In fact, astronomy related
observables such as the core radius $r_c,$  and the radial velocity dispersion
$\sigma,$ (assumed here to be given by the luminous content of the DSG's) are
shown to be free of divergence.

Note that the theoretical estimates  might also suffer from
an insufficient treatment of exchange terms in this mean field theory, but
exchange potentials are unlikely to reach astronomic  ranges.

We have seen, towards the end of Section II, that the theory provides
values of the ratio, ${\cal R}=\sigma^2/(G\, \rho_0\, r_c^2)$, that are almost
parameter independent, in particular from the temperature, and in agreement with
the King radius. This may lead to some trust in the theoretical treatment.

Obviously, the particle mass  $m$ is not really a parameter. Rather, it
should emerge as the same physical result for all observations. Our theory
easily allows, see Eq.(\ref{masfrmulb}), to estimate  $m$ from the two
traditional observables, $\sigma$ and $r_c$.  It will be noted that the
``slightly debatable'' functions, ${\cal F}(\eta)$ and ${\cal G}(\eta)$,
contained in Eq.(\ref{masfrmul}), have values not far from  ${\cal O}(1)$,
and vary smoothly only. With the former and the latter functions taken to
powers $1/8$ and $1/2$, respectively, their influence remains small. 

All being considered, our results can be expressed in terms of two
different fermionic mass scales, depending on the physical interpretation
of the stage at which the fermionic cloud is considered to be in a
non-relativistic thermal equilibrium. If one assumes that thermal equilibrium
describes the present observable stage of the DSG's, then one obtains a
fermion mass estimate of $mc^2 = 275 \pm 30\ eV.$  If one rather
assumes that the equilibrium corresponds to an initial stage only, followed
by a collision-less and dissipation-less evolution, one is led to an estimated
lower bound, $550\, eV \lesssim m\, c^2,$ according to the reasoning of
Refs. \cite{Boyarsky:2008ju,Shao:2012cg}. The compatibility, or possible
tension, of such determinations with other sources of astronomic observables
is beyond the focus of the present work. But it obviously deserves to be
studied.

As a final remark on our theory, we have not been able to design a precise way
to deduce a dark cloud temperature from astronomic  data, because our main
result, Eq.(\ref{masfrmulb}), does not seem to make the elementary mass to
depend on the ``$\eta$ parameter'' strongly enough. This problem will remain
under our consideration.

\section*{Acknowledgements}
We thank Patrick Valageas for remarks and suggestions on the manuscript.

\end{document}